# Network analysis identifies weak and strong links in a metapopulation system

Alejandro F. Rozenfeld[1], Sophie Arnaud-Haond[2,4], Emilio Hernández-García[3], Víctor M. Eguíluz[3], Ester A. Serrão[2] and Carlos M. Duarte[1]

1 IMEDEA (CSIC-UIB), Instituto Mediterráneo de Estudios Avanzados, C/ Miquel Marqués 21, 07190 Esporles, Mallorca, Spain.

2 CCMAR, CIMAR-Laboratório Associado, Universidade do Algarve, Gambelas, 8005-139, Faro, Portugal

3 IFISC, Instituto de Física Interdisciplinar y Sistemas Complejos (CSIC-UIB), Campus Universitat de les Illes Balears, E-07122 Palma de Mallorca, Spain.

4 IFREMER, Centre de Brest, BP70, 29280 Plouzané, France

**Corresponding Author :** Alejandro Rozenfeld, alex@ifisc.uib.es, Instituto Mediterráneo de Estudios Avanzados, C/ Miquel Marqués 21, 07190 Esporles, Mallorca, Spain.

**Manuscript information:** 15 pages, 6 figures, 2 tables.

**Abstract**

The identification of key populations shaping the structure and connectivity of metapopulation systems is a major challenge in population ecology. The use of molecular markers in the theoretical framework of population genetics has allowed great advances in this field, but the prime question of quantifying the role of each population in the system remains unresolved. Furthermore, the use and interpretation of classical methods are still bounded by the need for *a priori* information and underlying assumptions that are seldom respected in natural systems. Network theory was applied to map the genetic structure in a metapopulation system using microsatellite data from populations of a threatened seagrass, *Posidonia oceanica*, across its whole geographical range. The network approach, free from *a priori* assumptions and of usual underlying hypothesis required for the interpretation of classical analysis, allows both the straightforward characterization of hierarchical population structure and the detection of populations acting as hubs critical for relaying gene flow or sustaining the metapopulation system. This development opens major perspectives in ecology and evolution in general, particularly in areas such as conservation biology and epidemiology, where targeting specific populations is crucial.

Understanding the connectivity between components of a metapopulation system and their role as weak or strong links remains a major challenge of population ecology (1-3). Advances in molecular biology fostered the use of indirect approaches to understand metapopulation structure, based on describing the distribution of gene variants (alleles) in space within the theoretical framework of population genetics (4-7). Yet, the premises of the classical Wright-Fisher model (4, 6), such as "migration-drift" and "mutation-drift" equilibrium (8), "equal population sizes" or symmetrical rate migration among populations, are often violated in real metapopulation systems. Threatened or pathogen species, for example, are precisely studied for their state of demographic disequilibrium due to decline and local extinctions in the first case, or to their complex dynamics of local decline and sudden pandemic burst in the second. Furthermore, the underlying hypotheses of equal population size and symmetrical migration rates hamper the identification of putative population "hubs" centralizing migration pathways or acting as sources in a metapopulation system, which is a central issue in ecology in general, and in conservation biology or epidemiology in particular. Finally, complementary methods of genetic structure analyses, such as hierarchical AMOVA and coalescent methods rely on *a priori information* (or *prior*s) as to the clustering or demographic state of populations, requiring either subjective assumptions or the availability of reliable demographic, historical or ecological information that are seldom available.

Network theory is emerging as a powerful tool to understand the behavior of complex systems composed of many interacting units (9-11). Although network theory has been applied to a broad array of problems (12-14), only recently has it been adapted to examining genetic relationships among populations or individuals (15, 16). Yet, relevant properties of networks, such as resistance (9) to perturbations (i.e. node paralysis or destruction), the ability to host coherent oscillations (17) or the predominant importance of nodes or cluster of nodes in maintaining the integrity of the system or relaying information through it can be deducted from the network topology and specific characteristics (10, 11). Here we apply network theory to population genetics data of a threatened species, the Mediterranean clonal seagrass *Posidonia oceanica*. We start by analyzing data at the Mediterranean scale, where clustering of the populations distributed in two basins connected by a narrow straight and that were almost isolated during the last glaciation, was rather obvious *a priori*. This particular geographical and historical context facilitated the classical AMOVA analysis and allows using this example to validate our network analysis and confirm its potential, without *a priori* knowledge or assumptions, to characterize population genetic structure and to identify populations that are critical to the dynamics and sustainability of the whole system. We then compared classical and network tools at the regional scale of the Spanish coasts, a more common case in population genetics where no strong expectations can be suggested as to the distribution of connectivity among populations or clusters of populations. The results open major perspectives in evolutionary ecology, and more specifically in conservation biology and epidemiology where the capacity to target populations requiring major efforts towards conservation or control is crucial.

**Results and Discussion**

We build networks of population connectivity for a system of 37 meadows of the marine plant *Posidonia oceanica*, sampled across its entire geographic range -the Mediterranean Sea-, by using seven microsatellite markers (18). The network was built by considering any pair of populations as linked when their genetic distance (Goldstein distance (19)) is smaller than a suitably chosen distance threshold (see Methods). We highlight these links as the relevant genetic relationships either at the Mediterranean (the full dataset) or at the regional (28 populations along Spanish coasts) scales.

The topology of the network obtained at the Mediterranean scale (Fig. 1) highlights, without any a priori geographical information being used, the historical cleavage between Eastern and Western basins (18) and the transitional position of the populations from the Siculo-Tunisian Strait (see Fig. 1). Besides this graphical representation, closer to reality than the usual binary trees, indices derived from the holistic analyses of network topology allow unraveling some dynamic properties in terms of gene flow through this network of populations. The average *clustering coefficient*, $<C>=0.96$, is significantly higher than the one expected after randomly rewiring the links ($<C_0>=0.76$ with $\sigma_0=0.02$, after 10000 randomizations) revealing the existence of clusters of populations more interconnected than expected by chance. The values of *betweenness centrality*, quantifying the relative importance of the meadows in relaying information flow through the network, immediately highlight a meadow in Sicily (present in 21% of all shortest paths among populations), together with another one in Cyprus (16%), as the main stepping-stones between the pairs of populations sampled in the Western and Eastern basins, respectively (Fig. 1 and Table 1). These results are therefore in agreement with the genetic structure revealed with classical population genetics analysis (Analysis of Molecular Variance "AMOVA"), revealing past vicariance (18) and a secondary contact zone in the Siculo-Tunisian Strait. The metapopulation structure, clustering and 'transition zones' derived from the network analysis arise without any *a priori* input on clustering as needed for AMOVA, and without using geographic information in the analysis of allelic richness previously performed to support the existence and localization of a contact zone (18). This example allowed us to test the accuracy of network analysis on a population genetics dataset by comparing its results with a well understood case, where the expected clustering of populations and pathways are rendered almost obvious by the geography (two clusters of marine populations split by land and communicating only through a narrow straight). This comparison has been a first demonstration of the reliability of network analysis .

.

We then examined a non-trivial case: 28 populations sampled along Spain´s continental coasts and in the Balearic islands, more extensively and homogeneously sampled than the rest of the Mediterranean (Table 1). Classical tools resulted in a matrix of pairwise genetic distances ($F_{ST}$ or Goldstein distance) showing significant differentiation among almost all meadows (except one pair) without clear relationship with geographic situation, no clear pattern of allelic richness, and a "comb like" topology in a UPGMA Tree, which forces dichotomous branching of the metapopulation network (Figure 4a, see Methods). These methods were unable to highlight neither any particularly central position for the populations analyzed nor the clustering of some subgroups that would have suggested preferential roads for gene flow, or a dominant role of some populations in the metapopulation system. On the contrary, network analyses of these populations (Figure 2, Table 2) revealed a centralized structure with particularly important roles for certain populations. The *degree distribution*, *P(k)*, i.e. the proportion of nodes with *k* connections to other nodes, decays rapidly for large *k* (Fig 3.a) and the six highest values are all observed in samples collected in the Balearic Islands (Fig 2, Table 2). The average *clustering coefficient* of $<C>=0.4$ was significantly higher than that obtained in the corresponding randomized networks ($<C_0>=0.13$ with $\sigma_0=0.05$ after 10000 realizations), whereas the local clustering decays as a function of the degree *k* (Fig 3.b) which indicates that the central core is substructured into a small set of hubs, with high connectivity and low clustering, linking groups of closely connected nodes (i.e. with high clustering). Examination of the relationship between the degree of a node and the average degree of the populations connected to it showed an abundance of links between highly connected and poorly connected nodes (Fig 3.c), a property termed *dissortativity,* present in many biological networks (20)**,** and confirms again a centralized topology. Observation of Fig. 2 indicates that seagrass populations along the Spanish continental coasts are genetically closer to Balearic populations than to geographically closer populations. The highest values of *betweenness centrality* (Table 2) are also attained at the Balearic populations, suggesting that the meadows of this region play or have played a central role in relaying gene flow at the scale of the Spanish coasts**.** Moreover, the *betweenness centrality* increases exponentially with the *connectivity degree k* (Fig. 3d). All these findings reveal a star-like structure where hubs are connected in cascade and the central core is the set of Balearic populations. A clear but more constrained perspective of this pattern is partly shown by the resulting Minimum Spanning Tree (MST) of populations (Fig. 4b, see Methods) which, when analyzed with the network index of *betweenness centrality* highlights three of the major hubs encountered on the network. Yet, some other populations identified on the network appear as poorly connected on the MST, as a consequence of being a more constrained method, which finds the minimal paths required to maintain connectivity but not all the important ones. This emphasizes again the advantage of the network illustration and analysis. The biological implication of these results is a great centrality of the Balearic Islands, acting or having acted as a hub for gene flow thorough the system.

Populations with high degree *k* might either be sources sustaining the system (i.e. spreading propagules), or sinks receiving gene flow from all the other populations, or both. The extremely low rate of sexual recruitment inferred in populations with low clonal diversity ($R$) renders those, if highly connected, much more likely to disperse than to receive. The presence in the Balearic Islands of the two populations with the lowest observed clonal diversity and the highest connectivity (Es Port, $R=0.1$; $k=10$; and Fornells $R=0.1$; $k=15$), likely representing populations supplying "genetic material" to neighbor populations, suggests again that the Balearic islands are a key region for the dynamics and connectivity of the metapopulation system at the scale of the Spanish coast. Furthermore, 8 among 16 continental populations show extreme low connectivity ($k=0$), thereby allowing identification of those least likely to be rescued by other populations if threatened. As in any genetic approach to metapopulation management, the role of currently observed connectivity in future population rescuing is more important if current connectivity is limited by dispersal ability rather than by competitive interactions that could change in the future in decaying populations. Additionally, given the particular millenary nature of *P. oceanica* clones, current genetic structure is likely to integrate patterns of gene flow over past centuries, and thus may not reflect present-day dynamics.

Both networks, that at the scale of the whole Mediterranean (Fig. 1) and that for the Spanish coasts (Fig. 2), presented "small world" properties (21), i.e. a diameter ($L=1.39$ and $L=1.63$ respectively) shorter than expected for random networks ($<L_0>=1.47$ with $\sigma_0=0.01$ and $<L_0>=2.53$ with $\sigma_0=0.15$ respectively, after 10000 randomizations) whereas their clustering was much higher (see numerical values above), suggesting a highly hierarchical substructure. This provides clear evidence for the appearance of "short-cuts" in gene flow at multiple geographical scales along the history of this species, indicating rare events of large scale dispersal having a significant impact on the genetic composition of populations. This result illustrates another benefit of holistic (i.e. taking into account all distances or relationships among agents) network approaches. In the absence of other data, the possible existence of sources tends to be indirectly inferred with classical tools by pointing the populations exhibiting the highest levels of genetic diversity. Here, network analyses revealed that, although sexual recruitment in the targeted populations is very low, they are –or have been- important as paths of gene flow in the system.

Our results demonstrate that network analyses provide a holistic and powerful approach to unravel genetic structure and connectivity at different spatial scales. First, they are free of *a priori* hypotheses about the clustering of populations or heavy underlying assumptions -such as Fisher-Wright equilibrium- that are required to run or interpret classical analyses, but are seldom respected in nature. They allow to accurately study gene flow in clonal organisms, by removing the prerequisite of genotypic frequency equilibrium required by assignment based methods. Second, they allow unraveling properties that could not be highlighted by classical methods alone. The use of specific network properties such as the *betweenness centrality* and the *degree distribution* allows to identify populations relaying gene flow, or acting as sources supplying the system, in addition to achieving a quantitative ranking of populations depending on their respective roles in the dynamics of the system. Third, network analysis tools provided graphical representations of the genetic relatedness between populations in a multidimensional space (15), free of some of the constraints (e.g. binary branching) compulsory in classical methods describing population relationships. Finally, network analyses are based on distances, which will allow future users to modulate and choose, among the wide panel of distances applicable to molecular data, the most appropriate ones for the particular questions addressed, for the life history traits of the model organism, and for the type of molecular markers used. Here we chose

Goldstein distance to study the integration of gene flow over a geological time scale with microsatellite data on a clonal seagrass presenting a pattern of strong genetic structure and ancient divergence. Depending on the time scale that is relevant for the questions addressed, one may choose to use other markers and/or other distances. Additionally, comparison of networks obtained with different markers and distances for the same system may allow inferences as to the evolution of gene flow and connectivity through time. To conclude, adressing gene flow using network tools may prove a ground-breaking milestone in critical areas such as conservation biology, dealing with threatened or invasive species, and epidemiology, where the definition of target populations to be conserved or eradicated is of crucial importance.

**Materials and Methods**

**Molecular data.** About 40 *Posidonia* shoots collected at each of the 37 sampled populations (Fig.1 and Table 1) were genotyped with a previously selected set of seven dinucleotide microsatellites (22) allowing the identification of clones (also called genets for clonal plants). Clonal diversity was estimated for each population as described in Arnaud-Haond et al. (22), and replicates of the same clone were excluded for the estimation of inter-population distances. The matrix of interpopulation distances was built using Goldstein metrics (19), thus taking into account the level of molecular divergence among alleles, besides the differences in allelic frequencies.

**Networks.** We first built a fully connected network with the 37 populations considered as nodes. Each link joining pairs of populations was labeled with the Goldstein distance among them. We then removed links from this network of genetic similarity, starting from the one with the largest genetic distance and following in decreasing order, until the network reaches the percolation point (23, 24), beyond which it loses its integrity and fragments into small clusters. This means that gene flow across the whole system is disabled if connections at a distance smaller than this critical one, *Dp*, are removed. The precise location of this percolation point is made with the standard methodology adequate for finite systems (23, 24), i.e., by calculating the average size of the clusters excluding the largest one, $\langle S \rangle^* = \frac{1}{N} \sum_{s<S_{max}} s^2 n_s$, as a function of the last distance value removed, *thr*, and identifying the critical distance with the one at which <S>* has a maximum. *N* is the total number of nodes not included in the largest cluster and $n_s$ is the number of clusters containing *s* nodes. Here we find *Dp*=91, as shown in Fig. 5.

Once the network at percolation point is obtained, we analyzed its topology and characteristics (See Fig.1 and Table 1), and interpret those biologically. The first column in Table 1 contains also the estimated clonal diversity R of the different populations, defined as the proportion of different genotypes found with respect to the total number of collected shoots.

At the Spanish coasts scale, no percolation point is found using the above procedure, meaning that the genetic structure in this area is rather different from the one at the whole Mediterranean scale. To construct a useful network representation of the meadows genetic similarity, the following alternative process was applied in order to determine a relevant distance threshold, *thr*, above which links are discarded. At a very low threshold (*thr*=16, see Movie 1 provided as Supplementary Online Information) only the inner part of a central core, constituted by some meadows from the Balearic

Islands, is connected. As the threshold is increased new meadows (from the central Spanish coast) become connected (*thr*=20). Beyond that value, more peripheral meadows are connected from the northern and southern Spanish coasts. The geographical extension of the connected cluster (Fig. 6) grows with the distance threshold and an important jump occurs at *thr*=22, when the northern and southern coasts get connected for the first time. Further distance threshold increase does not contribute to geographical extension. Therefore, we find the value *thr*=22 and the resulting network as appropriate for topological characterization, since at this point the network contains a rich mixture of strong and weak links spanning all the available geographic scales within the Mediterranean Spanish coasts.

**Estimates of global and local properties of the network.** The degree $k_i$ of a given node $i$ is the number of other nodes linked to it (i.e., the number of neighbor nodes). The *distribution P(k)* gives the proportion of nodes in the network having degree $k$.

We denote by $E_i$ the number of links existing among the neighbors of node $i$. This quantity takes values between 0 and $E_i^{(\max)} = \frac{k_i(k_i-1)}{2}$, which is the case of a fully connected neighborhood. The clustering coefficient $C_i$ of node $i$ is defined as:

$$C_i = \frac{E_i}{E_i^{(\max)}} = \frac{2E_i}{k_i(k_i-1)}$$

The clustering coefficient of the whole network <*C*> is defined as the average of all individual clustering coefficients in the system. The degree dependent clustering *C(k)* is obtained after averaging $C_i$ for nodes with degree $k$.

Real networks exhibit correlations among their nodes (20, 25-30) that play an important role in the characterization of the network topology. Those node correlations are furthermore essential to understand the dynamical aspects such as spreading of information or their robustness against targeted or random removal of their elements. In social networks, nodes having many connections tend to be connected with other highly connected nodes. This characteristic is usually referred to as *assortativity*, or *assortative mixing*. On the other hand, technological and biological networks show rather the property that nodes having high degrees are preferably connected with nodes having low degrees, a property referred to as *dissortativity*. Assortativity is usually studied by determining the properties of the average degree $<k_{nn}>$ of neighbors of a node as a function of its degree $k$ (20, 29, 31). If this function is increasing, the network is assortative, since it shows that nodes of high degree connect, on average, to nodes of high degree. Alternatively, if the function is decreasing, as in our present case, the network is dissortative, as nodes of high degree tend to connect to nodes of lower degree. In this last case, the nodes with high degree are therefore central hubs ensuring the connection of the whole system.

The betweenness centrality (32) of node $i$, $bc(i)$, counts the fraction of shortest paths between pairs of nodes which pass through node $i$. Let $\sigma_{st}$ denote the number of shortest paths connecting nodes $s$ and $t$ and $\sigma_{st}(i)$ the number of those passing through the node $i$. Then,

$$bc(i) = \sum_{s \neq t \neq i} \frac{\sigma_{st}(i)}{\sigma_{st}}.$$

The degree-dependent betweeness, $bc(k)$, is the average betweeness value of nodes having degree $k$.

**Minimum Spanning Tree.** Given a connected, undirected graph, a spanning tree of that graph is a subgraph without cycles which connects all the vertices together. A single graph can have many different spanning trees. Provided each edge is labeled with a cost (in our analysis the genetic distance among the connected populations) each spanning tree can be characterized by the sum of the cost of its edges. A minimum spanning tree is then a spanning tree with minimal total cost. A minimum spanning tree is in fact the minimum-cost subgraph connecting all vertices, since subgraphs containing cycles necessarily have more total cost. Figure 4 shows the minimum spanning tree for the Spanish meadows. The star-like structure centered at Balearic populations is evident, although the restriction of being a tree prevents some of the well connected populations of the network approach to be identified here.

We acknowledge financial support from the Spanish MEC (Spain) and FEDER through project FISICOS (FIS2007-60327), the Portuguese FCT and FEDER through project NETWORK(POCI/MAR/57342/2004) a postdoctoral fellowship (SAH), the BBVA Foundation (Spain), and the European Commission through the NEST-Complexity project EDEN (043251). We wish to thank François Bonhomme and Pierre Boudry for useful discussions, and two anonymous referees for constructive criticisms and suggestions that helped improving a previous manuscript.

**References.**

1. Haydon, D. T., Cleaveland, S., Taylor, L. H. & Laurenson, M. K. (2002) Identifying reservoirs of infection: A conceptual and practical challenge. *Emerging Infectious Diseases* **8,** 1468-1473.
2. Travis, J. M. J. & Park, K. J. (2004) Spatial structure and the control of invasive alien species. *Animal Conservation* **7,** 321-330.
3. Williams, J. C., ReVelle, C. S. & Levin, S. A. (2004) Using mathematical optimization models to design nature reserves. *Frontiers in Ecology and the Environment* **2,** 98-105.

4. Fisher, R. A. (1930) *The genetical theory of natural selection* (Clarendon press.
5. Haldane, J. (1932) *The Causes of Evolution* (Longmans Green.
6. Wright, S. (1931) Evolution in Mendelian populations. *Genetics* **16,** 97-159.
7. Wright, S. (1943) Isolation by distance. *Genetics* **28,** 114-138.
8. Hey, J. & Machado, C. A. (2003) The study of structured populations - New hope for a difficult and divided science. *Nat. Rev. Genet.* **4,** 535-543.
9. Albert, R., Jeong, H. & Barabasi, A. L. (2000) Error and attack tolerance of complex networks. *Nature* **406,** 378-382.
10. Amaral, L. A. N., Scala, A., Barthelemy, M. & Stanley, H. E. (2000) Classes of small-world networks. *Proc. Natl. Acad. Sci. U. S. A.* **97,** 11149-11152.
11. Strogatz, S. H. (2001) Exploring complex networks. *Nature* **410,** 268-276.
12. May, R. M. (2006) Network structure and the biology of populations. *Trends Ecol. Evol.* **21,** 394-399.
13. Proulx, S. R., Promislow, D. E. L. & Phillips, P. C. (2005) Network thinking in ecology and evolution. *Trends Ecol. Evol.* **20,** 345-353.
14. Watts, D. J. (1999) Networks, dynamics, and the small-world phenomenon. *American Journal of Sociology* **105,** 493-527.
15. Dyer, R. J. & Nason, J. D. (2004) Population Graphs: the graph theoretic shape of genetic structure. *Mol. Ecol.* **13,** 1713-1727.
16. Rozenfeld, A. F., Arnaud-Haond, S., Hernandez-Garcia, E., Eguiluz, V. M., Serrao, E. A. & Duarte, C. M. (2007) Spectrum of genetic diversity and networks of clonal populations. *Journal of the Royal Society Interface* **4,** 1093.
17. Lago-Fernandez, L. F., Huerta, R., Corbacho, F. & Siguenza, J. A. (2000) Fast response and temporal coherent oscillations in small-world networks. *Phys. Rev. Lett.* **84,** 2758-2761.
18. Arnaud-Haond, S., Migliaccio, M., Diaz-Almela, E., Teixeira, S., van de Vliet, M. S., Alberto, F., Procaccini, G., Duarte, C. M. & Serrao, E. A. (2007) Vicariance patterns in the Mediterranean Sea: east-west cleavage and low dispersal in the endemic seagrass Posidonia oceanica. *Journal of Biogeography* **34,** 963-976.
19. Goldstein, D. B., Linares, A. R., Cavalli-Sforza, L. L. & Feldman, M. W. (1995) Genetic Absolute Dating Based on Microsatellites and the Origin of Modern Humans. *Proc. Natl. Acad. Sci. U. S. A.* **92,** 6723-6727.
20. Newman, M. E. J. (2002) Assortative mixing in networks. *Phys. Rev. Lett.* **89,** 208701.
21. Watts, D. J. & Strogatz, S. H. (1998) Collective dynamics of 'small-world' networks. *Nature* **393,** 440-442.
22. Arnaud-Haond, S., Alberto, F., Teixeira, S., Procaccini, G., Serrao, E. A. & Duarte, C. M. (2005) Assessing genetic diversity in clonal organisms: Low diversity or low resolution? Combining power and cost efficiency in selecting markers. *J. Hered.* **96,** 434-440.
23. Grimmett, G. (1999) *Percolation* (Springer-Verlag, Berlin).
24. Stauffer, D. & Aharony, A. (1994) *Introduction to Percolation Theory* (Taylor & Francis, CRC Press, London).
25. Barrat, A. & Pastor-Satorras, R. (2005) Rate equation approach for correlations in growing network models. *Phys. Rev. E* **71,** 036127.
26. Boguñá, M. & Pastor-Satorras, R. (2003) Class of correlated random networks with hidden variables. *Phys. Rev. E* **68,** 036112.
27. Maslov, S. & Sneppen, K. (2002) Specificity and stability in topology of protein networks. *Science* **296,** 910-913.

28. Newman, M. E. J. (2003) Mixing patterns in networks. *Phys. Rev. E* **67,** 026126.
29. Pastor-Satorras, R., Vazquez, A. & Vespignani, A. (2001) Dynamical and correlation properties of the Internet. *Phys. Rev. Lett.* **87,** 258701.
30. Vazquez, A., Pastor-Satorras, R. & Vespignani, A. (2002) Large-scale topological and dynamical properties of the Internet. *Phys. Rev. E* **65,** 066130.
31. Lee, S. H., Kim, P. J. & Jeong, H. (2006) Statistical properties of sampled networks. *Phys. Rev. E* **73,** 016102
32. Freeman, L. C. (1977) Set of Measures of Centrality Based on Betweenness. *Sociometry* **40,** 35-41.

**Figure Captions:**

**Figure 1:** The network of Mediterranean meadows in which only links with Goldstein distances smaller than the percolation distance $Dp=91$ (see Fig. 5) are present. Nodes representing populations are roughly arranged according to their geographic origin. The precise geographic locations are indicated as diamonds in the background map. One can identify two clusters of meadows, corresponding to the Mediterranean basins (east and west), separated by the Siculo-Tunisian Strait. The size of each node indicates its *betweenness centrality* (*i.e.* the proportion of all shortest paths getting through the node).

**Figure 2:** The network constructed for the Spanish meadows with the "geographic threshold" criterion (see Fig. 6). Nodes are shown at the populations' geographic locations. Node sizes characterize their *betweenness centrality* (*i.e.* the proportion of all shortest paths getting through the node).

**Figure 3:** Main topological properties found by analysing the structure of the network of meadows at the Spanish basin scale (Fig. 2). (a) The complementary cumulative *degree distribution* $P(degree>k)$, (b) the local *clustering* $C(k)$, (c) the average degree $\langle k_{nn}(k) \rangle$ in the neighbourhood of a meadow with degree $k$, and (d) the *degree-dependent betweenness, $bc(k)$*, as a function of the connectivity degree $k$.

**Figure 4:** (a) UPGMA tree based on Goldstein distances, displaying a quite uninformative comb-like structure (b) Minimum Spanning Tree (MST) based on Goldstein distance among Spanish meadows. This is the subgraph which connects the populations at the Spanish coast scale minimizing the total genetic distance along links, with the diameter of nodes illustrating their index of betweeness centrality according to the topology of the MST.

**Figure 5.** The average cluster size excluding the largest one, as a function of the imposed genetic threshold, at the whole Mediterranean scale. This identifies $Dp$=91 as the percolation threshold.

**Figure 6.** The maximal geographic distance connected (at the Spanish coasts scale) as a function of the imposed genetic distance threshold (*thr*). Above *thr*=22 the maximal geographic distance covered by connected populations nearly duplicates, and this value is chosen to construct the corresponding network.

**Table 1:** Local properties of the whole Mediterranean network for *thr=Dp*=91. Information is given for the *betweenness centrality* (bc) and *clustering* (C), as well as clonal diversity estimates (*R*) for each sample.

**Table 2:** Local properties of the network constructed with the Spanish meadows. Information is given for the *connectivity degree* (*k*), *betweenness centrality* (bc) and *clustering* (C), as well as clonal diversity estimates (*R*) for each sample.

**Supporting Online Information**

**Movie 1:** Sequence of networks for the Spanish populations obtained at successive values of the threshold genetic distance, *thr*, above which links are discarded.

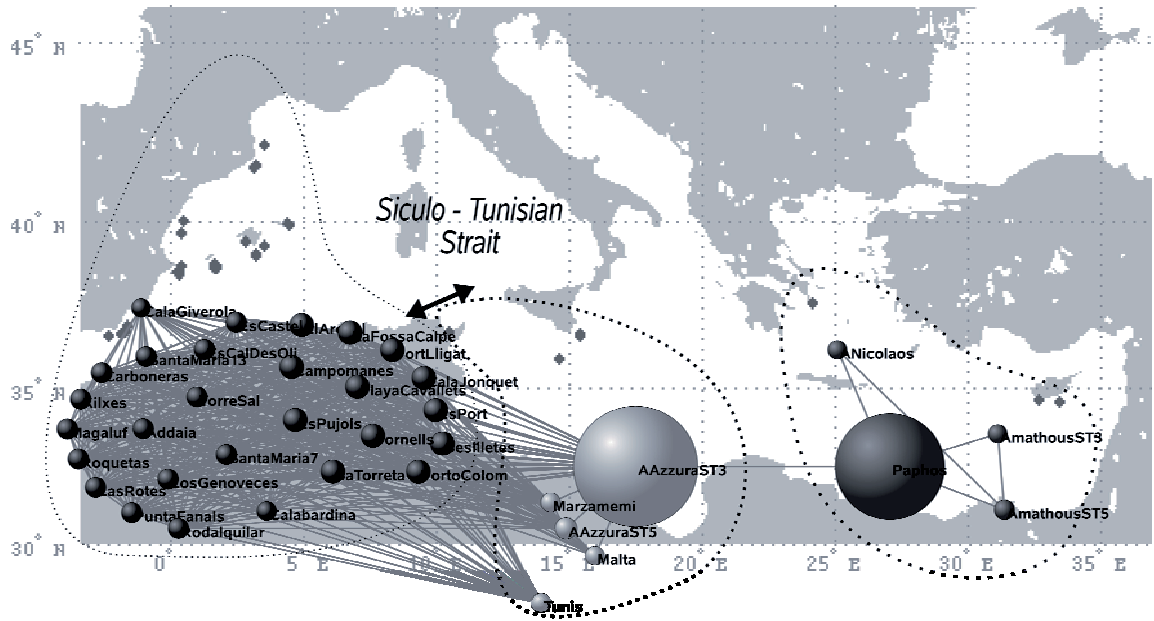

**Figure 1:** The network of Mediterranean meadows in which only links with Goldstein distances smaller than the percolation distance $Dp=91$ (see Fig. 5) are present. Nodes representing populations are roughly arranged according to their geographic origin. The precise geographic locations are indicated as diamonds in the background map. One can identify two clusters of meadows, corresponding to the Mediterranean basins (east and west), separated by the Siculo-Tunisian Strait. The size of each node indicates its *betweenness centrality* (*i.e.* the proportion of all shortest paths getting through the node).

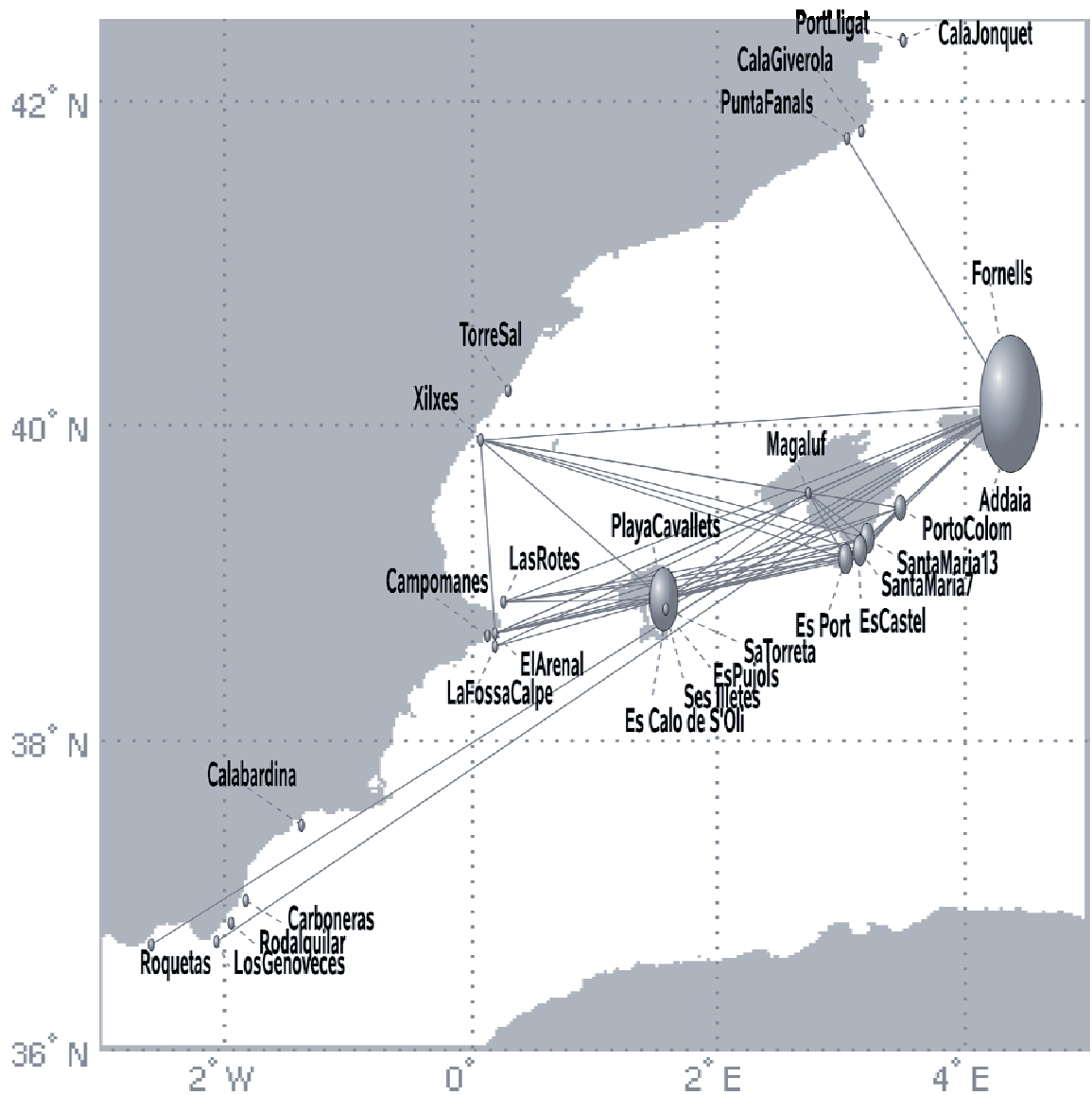

**Figure 2:** The network constructed for the Spanish meadows with the "geographic threshold" criterion (see Fig. 6). Nodes are shown at the populations' geographic locations. Node sizes characterize their *betweenness centrality* (*i.e.* the proportion of all shortest paths getting through the node).

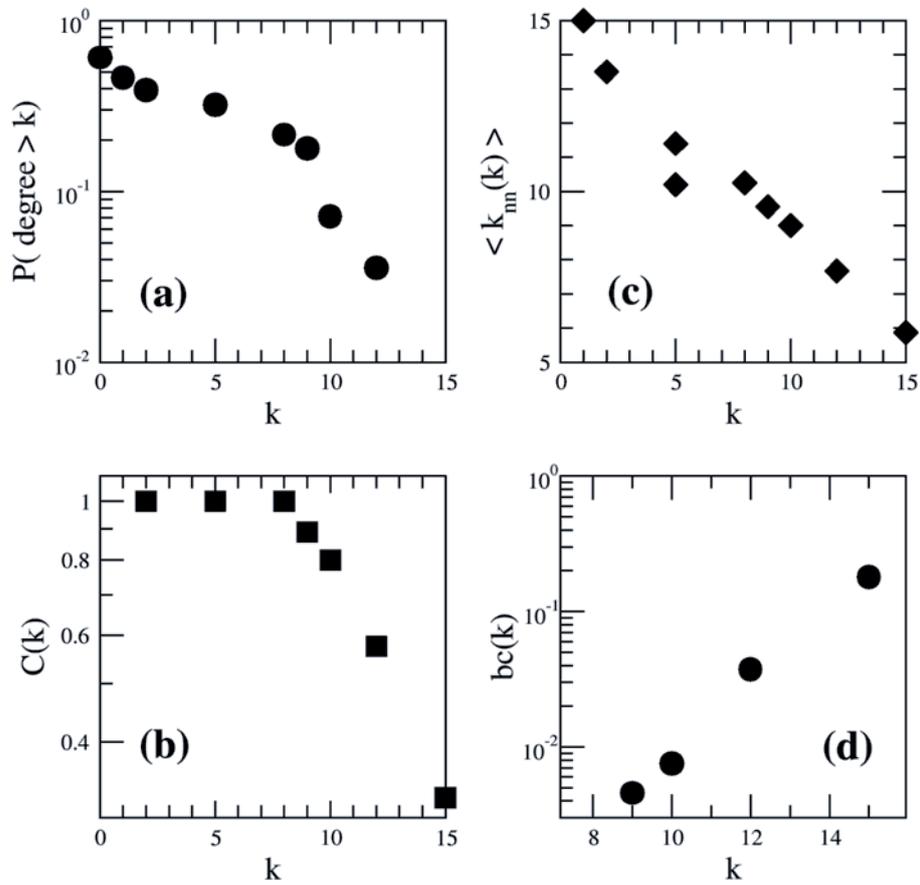

**Figure 3:** Main topological properties found by analysing the structure of the network of meadows at the Spanish basin scale (Fig. 2). (a) The complementary cumulative *degree distribution* P(degree>k), (b) the local *clustering C(k)*, (c) the average degree $<k_{nn}(k)>$ in the neighbourhood of a meadow with degree *k*, and (d) the *degree-dependent betweenness, bc(k),* as a function of the connectivity degree *k*.

a)

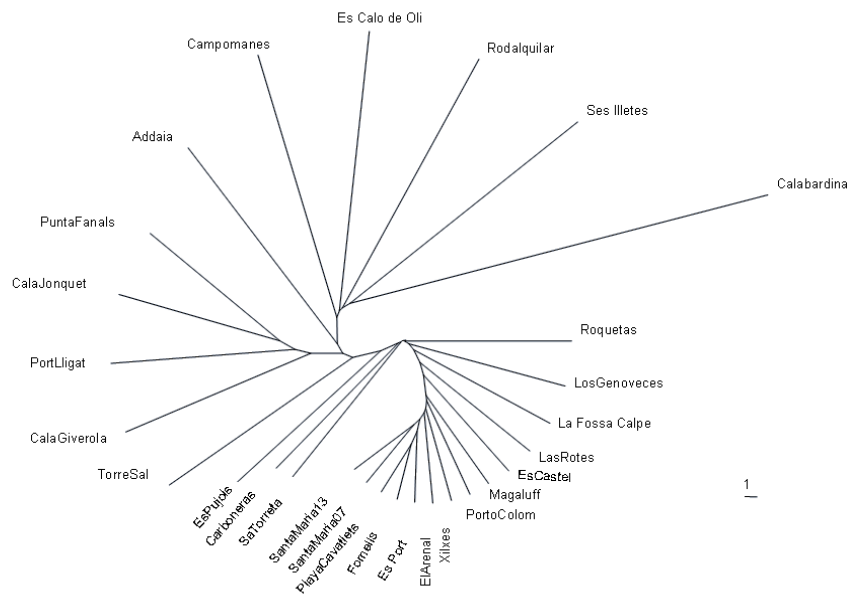

b)

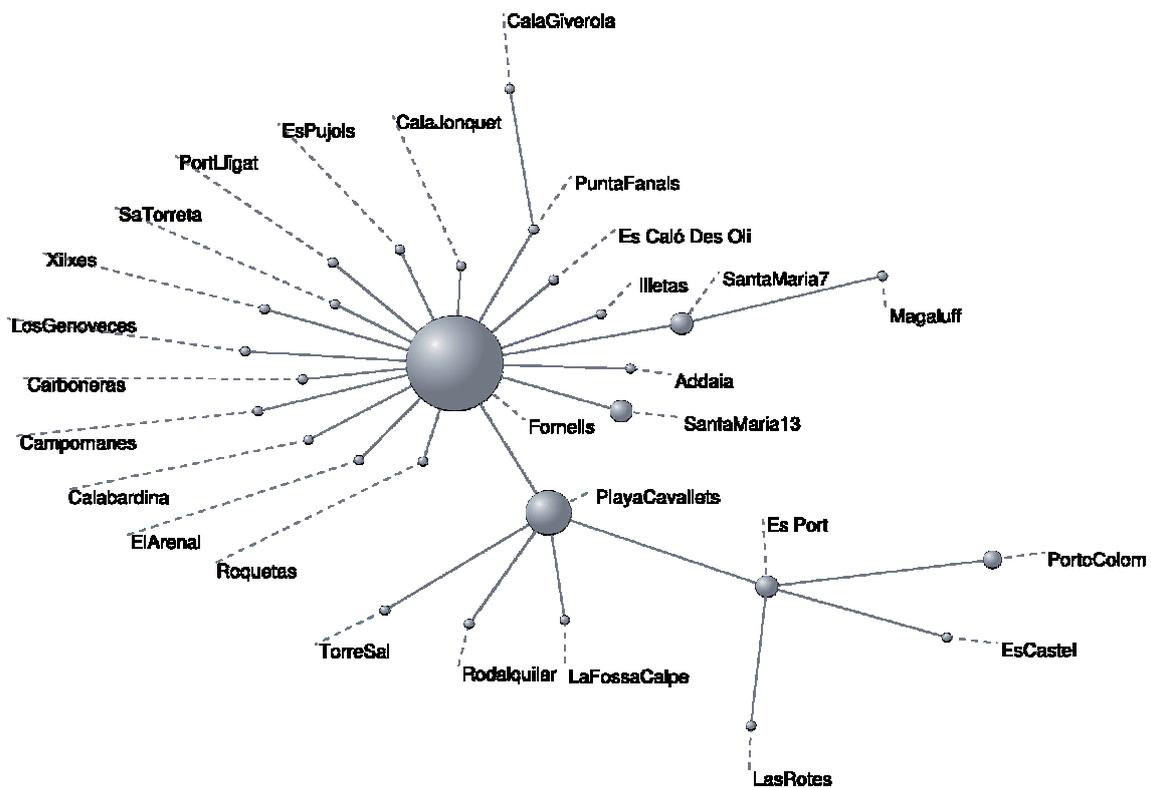

**Figure 4:** (a) UPGMA tree based on Goldstein distances, displaying a quite uninformative comb-like structure (b) Minimum Spanning Tree (MST) based on Goldstein distance among Spanish meadows. This is the subgraph which connects the populations at the Spanish coast scale minimizing the total genetic distance along links, with the diameter of nodes illustrating their index of betweeness centrality according to the topology of the MST.

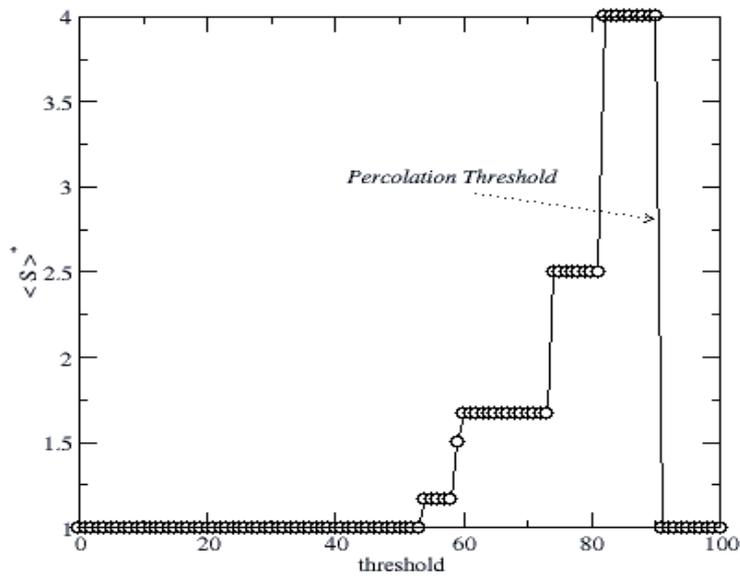

**Figure 5.** The average cluster size excluding the largest one, as a function of the imposed genetic threshold, at the whole Mediterranean scale. This identifies *Dp*=91 as the percolation threshold.

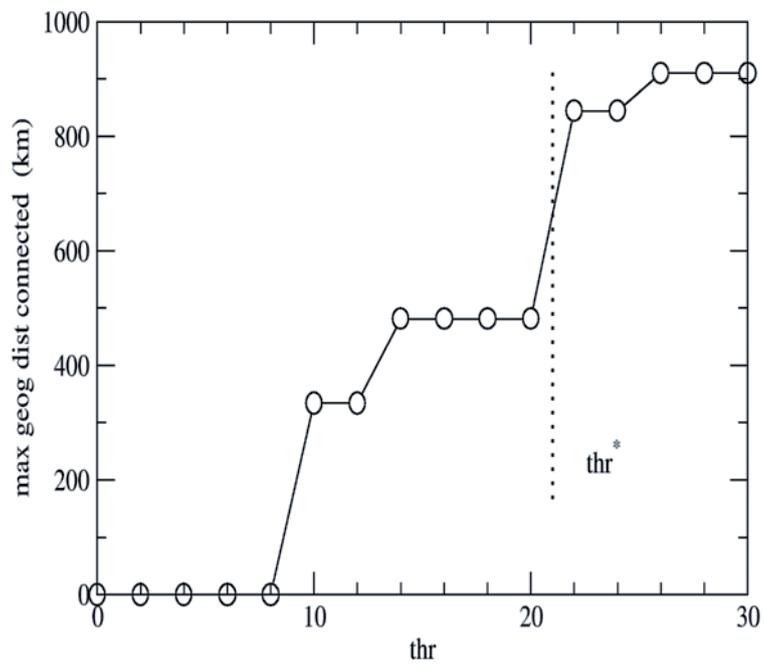

**Figure 6.** The maximal geographic distance connected (at the Spanish coasts scale) as a function of the imposed distance threshold (*thr*). Above *thr*=22 the maximal geographic distance covered by connected populations nearly duplicates, and this value is chosen to construct the corresponding network.

**Table 1:** Local properties of the whole Mediterranean network for *thr=Dp*=91. Information is given for the *betweenness centrality* (bc) and *clustering* (C), as well as clonal diversity estimates (*R*) for each sample.

| REGION | | Name | R | bc | C | REGION | Name | R | Bc | C |
|---|---|---|---|---|---|---|---|---|---|---|
| **SPANISH BALEARIC ISLANDS** | *Menorca* | Addaia | 0,67 | 0,0010 | 0,980 | **SPANISH IBERIAN PENINSULA** *(ORDERED FROM NORTH TO SOUTH)* | Cala Jonquet | 0,5 | 0,0031 | 0,946 |
| | | Fornells | 0,1 | 0,0031 | 0,946 | | Port Lligat | 0,28 | 0,0031 | 0,946 |
| | | | | | | | Cala Giverola | 0,43 | 0 | 0,997 |
| | *Mallorca* | Magaluf | 0,68 | 0,0010 | 0,981 | | Punta Fanals | 0,68 | 0,0010 | 0,981 |
| | | Porto Colom | 0,5 | 0,0031 | 0,946 | | Torre Sal | 0,5 | 0,0010 | 0,981 |
| | *Cabrera* | Es Castel | 0,1 | 0,0010 | 0,981 | | Xilxes | 0,35 | 0,0010 | 0,981 |
| | | Es Port | 0,1 | 0,0031 | 0,946 | | Las Rotes | 0,73 | 0,0010 | 0,981 |
| | | Santa Maria 13 | 0,56 | 0,0010 | 0,981 | | El Arenal | 0,86 | 0,0031 | 0,946 |
| | | Santa Maria 7 | 0,54 | 0,0010 | 0,981 | | Campomanes | 0,7 | 0,0031 | 0,946 |
| | *Ibiza* | Playa Cavallets | 0,73 | 0,0031 | 0,946 | | LaFossaCalpe | 0,77 | 0,0031 | 0,946 |
| | | | | | | | Calabardina | 0,88 | 0,0003 | 0,997 |
| | | Es Pujols | 0,67 | 0,0031 | 0,946 | | Carboneras | 0,32 | 0,0010 | 0,981 |
| | *Formentera* | EsCalo de S'Oli | 0,36 | 0,0010 | 0,981 | | Rodalquilar | 0,53 | 0,0010 | 0,981 |
| | | Ses Illetes | 0,6 | 0,0031 | 0,946 | | Los Genoveces | 0,34 | 0,0010 | 0,981 |
| | | Sa Torreta | 0,51 | 0,0031 | 0,946 | | Roquetas | 0,69 | 0,0010 | 0,981 |
| **CENTRAL BASIN** | *Sicily* | Tunis | 0,85 | 0 | 1 | **EAST BASIN** *Cyprus* | Amathous ST3 | 0,44 | 0 | 1 |
| | | Malta | 0,74 | 0 | 1 | | Amathous ST5 | 0,62 | 0,0008 | 0,667 |
| | | A. AzzuraST3 | 0,77 | 0,205 | 0,897 | | Paphos | 0,68 | 0,1579 | 0,333 |
| | | A. AzzuraST5 | 0,72 | 0,0017 | 0,963 | *Greece* | A. Nicolaos | 0,69 | 0 | 1 |
| | | Marzamemi | 0,81 | 0,0003 | 0,995 | | | | | |

**Table 2:** Local properties of the network constructed with the Spanish meadows. Information is given for the *connectivity degree* (*k*), *betweenness centrality* (bc) and *clustering* (*C*), as well as clonal diversity estimates (*R*) for each sample.

| REGION | | Name | R | k | bc | C | REGION | Name | R | k | bc | C |
|---|---|---|---|---|---|---|---|---|---|---|---|---|
| **SPANISH BALEARIC ISLANDS** | *Menorca* | Addaia | 0,67 | 0 | 0 | 0 | **SPANISH IBERIAN PENINSULA** *(ORDERED FROM NORTH TO SOUTH)* | Cala Jonquet | 0,5 | 0 | 0 | 0 |
| | | | | | | | | Port Lligat | 0,28 | 0 | 0 | 0 |
| | | Fornells | 0,1 | 15 | 0,180 | 0,32 | | | | | | |
| | | | | | | | | Cala Giverola | 0,43 | 0 | 0 | 0 |
| | | | | | | | | Punta Fanals | 0,68 | 1 | 0 | 0 |
| | *Mallorca* | Magaluf | 0,68 | 8 | 0 | 1 | | Torre Sal | 0,5 | 0 | 0 | 0 |
| | | Porto Colom | 0,5 | 9 | 0,0046 | 0,89 | | Xilxes | 0,35 | 8 | 0 | 1 |
| | | Es Castel | 0,1 | 5 | 0 | 1 | | Las Rotes | 0,73 | 5 | 0 | 1 |
| | *Cabrera* | Es Port | 0,1 | 10 | 0,0075 | 0,8 | | El Arenal | 0,86 | 8 | 0 | 1 |
| | | Santa Maria 13 | 0,56 | 10 | 0,0075 | 0,8 | | Campomanes | 0,7 | 0 | 0 | 0 |
| | | Santa Maria 7 | 0,54 | 10 | 0,0075 | 0,8 | | LaFossaCalpe | 0,77 | 2 | 0 | 1 |
| | *Ibiza* | Playa Cavallets | 0,73 | 12 | 0,0037 | 0,58 | | Calabardina | 0,88 | 0 | 0 | 0 |
| | | Es Pujols | 0,67 | 1 | 0 | 0 | | Carboneras | 0,32 | 0 | 0 | 0 |
| | *Formentera* | EsCalo de S'Oli | 0,36 | 0 | 0 | 0 | | Rodalquilar | 0,53 | 0 | 0 | 0 |
| | | Ses Illetes | 0,6 | 0 | 0 | 0 | | Los Genoveces | 0,34 | 1 | 0 | 0 |
| | | Sa Torreta | 0,51 | 2 | 0 | 1 | | Roquetas | 0,69 | 1 | 0 | 0 |